# Self-diffusion in two-dimensional hard ellipsoid suspensions

Zhongyu Zheng and Yilong Han*

Department of Physics, Hong Kong University of Science and Technology,
Clear Water Bay, Hong Kong, China

* Electronic address: yilong@ust.hk

**Abstract**

We studied the self-diffusion of colloidal ellipsoids in a monolayer near a flat wall by video microscopy. The image processing algorithm can track the positions and orientations of ellipsoids with sub-pixel resolution. The translational and rotational diffusions were measured in both the lab frame and the body frame along the long and short axes. The long-time and short-time diffusion coefficients of translational and rotational motions were measured as functions of the particle concentration. We observed sub-diffusive behavior in the intermediate time regime due to the caging of neighboring particles. Both the beginning and the ending times of the intermediate regime exhibit power-law dependence on concentration. The long-time and short-time diffusion anisotropies change non-monotonically with concentration and reach minima in the semi-dilute regime because the motions along long axes are caged at lower concentrations than the motions along short axes. The effective diffusion coefficients change with time $t$ as a linear function of $(\ln t)/t$ for the translational and rotational diffusions at various particle densities. This indicates that their relaxation functions decay according to $1/t$ which provides new challenges in theory. The effects of coupling between rotational and translational Brownian motions were demonstrated and the two time scales corresponding to anisotropic particle shape and anisotropic neighboring environment were measured.

## I. Introduction

Particle diffusions at interfaces are of great importance in many biological and industrial processes, such as coatings [1, 2], catalysis [3], Pickering emulsions [4] and protein diffusion in membranes [5]. In most real processes, particles are not spheres.

As a first-order approximation of non-spherical particles, rod-like particles or ellipsoids exhibit rich Brownian dynamics and phase behaviors. In the past, the aspect ratio and concentration effects of anisotropic particles have been measured using various experimental methods, such as depolarized dynamic light scattering [6], fluorescence recovery after photobleaching (FRAP) [7, 8], transient electric birefringence (TEB) [9, 10] and fluorescence correlation spectroscopy [11, 12] in three-dimensional (3D) suspensions. In two-dimensional (2D) suspensions, the aspect ratio and confinement effects have been investigated in the diffusions of single ellipsoids [13, 14], carbon nanotubes [15, 16] and actin filaments [17] between two walls. However, our understanding of the impact of concentration on the diffusions of anisotropic particles in 2D suspensions is far from sufficient, especially in the short-time regime. To our knowledge, there is no systematic experimental study on the concentration dependence of the diffusion of non-spherical particles in 2D.

In this paper, we investigate the Brownian motions of ellipsoids in a monolayer near a wall at different concentrations by digital video microscopy. Unlike other experimental techniques such as the dynamic light scattering [18-20] and nuclear magnetic resonance (NMR) [20, 21] which only yield information about ensemble averaged quantities, video microscopy reveals detailed single-particle dynamics. This technique has been used to measure the diffusions of ellipsoids in 2D [13, 14] and nanorods [15, 16, 22] at the infinite dilute limit. Here we developed an image analysis algorithm to track ellipsoids in a dense monolayer and studied their diffusions.

The self-diffusion coefficient $D$ is directly related to the drag coefficient $\gamma$ via the Einstein relation

$$D = k_B T / \gamma , \qquad (1)$$

where $k_B$ is the Boltzmann constant and $T$ is the temperature. $\gamma$ is sensitive to particle concentration, particle size, shape and interaction. The anisotropic particle shape leads to anisotropic drag coefficient. For a prolate ellipsoid with a long axis of length $2a$ and two short axes of length $2b$, the translational diffusion is described by diffusion coefficients $D_\parallel = k_B T / \gamma_\parallel$ along the long axis and $D_\perp = k_B T / \gamma_\perp$ along the short axes. The rotational diffusion coefficient about the short axis is $D_\theta = k_B T / \gamma_\theta$ . The translational and rotational drag coefficients are

$$\gamma_{\|,\perp} = 6\pi\eta b G_{\|,\perp}, \tag{2a}$$

$$\gamma_\theta = 6\eta V G_\theta, \tag{2b}$$

where the volume $V = 4\pi ab^2/3$ and $G$ is the geometric factor characterizing the amount of deviation of the ellipsoid from a sphere. The geometric factors for a prolate ellipsoid diffusing in an unbounded static 3D fluid were analytically derived by F. Perrin, the son of J. Perrin [23-25]:

$$G_a = \frac{8}{3}\left[\frac{2p}{1-p^2} + \frac{2p^2-1}{(p^2-1)^{3/2}}\ln\left(\frac{p+\sqrt{p^2-1}}{p-\sqrt{p^2-1}}\right)\right]^{-1}, \tag{3a}$$

$$G_b = \frac{8}{3}\left[\frac{p}{p^2-1} + \frac{2p^2-3}{(p^2-1)^{3/2}}\ln(p+\sqrt{p^2-1})\right]^{-1}, \tag{3b}$$

$$G_\theta = \frac{2}{3}\frac{(p^4-1)}{p}\left[\frac{2p^2-1}{\sqrt{p^2-1}}\ln(p+\sqrt{p^2-1}) - p\right]^{-1}, \tag{3c}$$

where $p = a/b$ is the aspect ratio.

As the area fraction, $\phi$, of the particles increases, $\gamma(\phi)$ increases and $D(\phi)$ decreases from $D(0)$ at infinite dilution to 0 at the solid phase. The diffusion of a particle at finite concentration is characterized by three time regimes, $\tau_B \ll t \ll \tau_R$, $t \sim \tau_R$ and $t \gg \tau_R$, due to the cage effect from its neighboring particles [26]. The Brownian relaxation time $\tau_B = m/(6\pi\eta R)$, where $m$ is the particle mass, $\eta$ is the fluid viscosity and $R$ is the radius of the particle. It characterizes the very short time scale from pre-Brownian ballistic motion to Brownian diffusive motion. For our micrometer-sized ellipsoids, $\tau_B \sim 10^{-6}$ s. $\tau_R \approx R^2/(4D(0)) \approx ab/(4D(0)) = 30$ s is the time for a particle to diffuse a distance comparable to its size. It is an estimate of the time scale at which the cage effect and direct interactions with neighbors (mainly excluded volume effect for hard particles) become important. In the Brownian short-time regime $\tau_B \ll t \ll \tau_R$, the motion is diffusive. The configuration of the suspension does not change significantly, thus the inter-particle interactions are the indirect hydrodynamic interactions transmitted by the fluids. At the intermediate time scale $t \sim \tau_R$, the particle is caged by neighbors and experiences sub-diffusion. At long time $t \gg \tau_R$, the particle diffuses out of cages and the motions become diffusive again, albeit at a slower rate.

The three time regimes can be resolved from the mean-square displacement (MSD), $MSD = \langle \Delta r^2(t) \rangle$, where $\Delta r$ is the displacement of a particle during time $t$ and $\langle \rangle$ is the ensemble average. MSD increases linearly in the short-time regime ($\tau_B \ll t \ll \tau_R$), curves downward in the intermediate time regime ($t \sim \tau_R$) due to the cage effect, and returns to a straight line but with a smaller slope in the long-time regime ($t \gg \tau_R$). The slope of the *MSD* is proportional to the diffusion coefficient $D(t) = \Delta MSD / (2d\Delta t)$ where $d$ is the dimension of the space. In the long- and short-time regimes, the motions are diffusive and the *MSD*s have constant slopes. We measured the diffusion coefficients ($D_T^L$, $D_\theta^L$, $D_T^S$, $D_\theta^S$) of ellipsoids in the lab frame, where the superscripts, *L* and *S*, represent long-time and short-time regimes and the subscripts, *T* and $\theta$, represent translation and rotation respectively. We further projected the trajectories into each particle's body frame and measured the diffusion coefficients ($D_\parallel^L$, $D_\perp^L$, $D_\parallel^S$, $D_\perp^S$) along the long ($\parallel$) and short ($\perp$) axes. The diffusion anisotropy as characterized by $D_\parallel/D_\perp$ and the translation-rotation coupling were measured and discussed. Although the *MSD* in each degree of freedom is not linear in the intermediate time regime, its corresponding effective diffusion coefficient $\tilde{D}(t) = MSD/(2t)$ can be expressed as a linear function of $(\ln t)/t$ for spheres in 2D. Interestingly we found that such a linear relation still holds for ellipsoids, which indicates that the relaxation functions are inversely proportional to time and the diffusions are dominated by direct binary collisions in our measured density regime.

**II. Materials and Methods**

The polymethyl methacrylate (PMMA) ellipsoids were synthesized through the method described in ref. [27]. Briefly, we added PMMA spheres into an aqueous polyvinyl alcohol (PVA) solution in a Petri dish. After water was evaporated at room temperature, the PVA film was stretched at 130ºC which is above the glass transition temperatures of PVA ($T_g$ = 85ºC) and PMMA ($T_g$ = 105ºC) so that the PMMA spheres could be deformed into ellipsoids. After cooling to room temperature, the PVA was dissolved in water and ellipsoids were obtained. The ellipsoids had 5.8% polydispersity with semi-long axis $a$ = 5.9 ± 0.34 μm and semi-short axes $b = c$ = 0.65 ± 0.04 μm. The cleaned aqueous suspensions of ellipsoids were stabilized with 7 mM sodium dodecyl sulfate (SDS). The ionic strength was more than 0.1 mM and the corresponding Debye screening length was less than 30 nm. Hence the ellipsoids can be considered as hard particles and double layers of counterions did not significantly affect their diffusions. The aqueous suspension of ellipsoids was enclosed in a 12 ×15× 0.1 mm$^3$ glass cell and sealed with epoxy glue. The bottom glass surface was coated with polydimethylsiloxane (PDMS) in order to avoid particles sticking to it. The ellipsoids were heavy enough to sediment on the bottom and the height fluctuation was very small. Therefore, the ellipsoids diffused in the two-dimensional horizontal plane near the bottom. The drag force imposed by the upper glass ceiling can be neglected [28] because the distance between the upper wall and ellipsoids was about 200$b$. The diffusions of ellipsoids were measured after the samples had been equilibrated for 6 hours. The motions of the ellipsoids were observed using an optical microscope with a 20× objective and recorded by a charge coupled device (CCD) camera at 15 frames per second for 60 minutes at each particle concentration. This frame rate and the measurement time enabled us to explore all the three time regimes for the ellipsoids. At different concentrations, several tens to several thousands of ellipsoids were uniformly distributed within the 640×480 pixels (1 pixel = 0.33 μm) field of view (see Fig. 1(a)). No drift flow was observed in the suspension during the one-hour experimental period at each concentration. All experiments were carried out at a room temperature of 23°C.

We developed an image processing algorithm to track multiple ellipsoids in bright-field images as shown in Fig. 1. The algorithm performed the following operations. First, it applied a spatial band-pass filter to remove the noises in the image [29], and then used the standard sphere-tracking algorithm in ref. [29] to identify the candidate positions of ellipsoids. However, one ellipsoid might be incorrectly tracked as multiple spheres, i.e. multiple candidate positions of the center of mass. So next, our algorithm removed the wrong candidate positions and roughly estimated the true position of each ellipsoid in a 4$a$ × 4$a$ area centered around each position. We created 18 4$a$ × 4$a$ masks; each mask had an ellipse at the center with 18 different orientations: 0º, 10º, 20º,…, 170º. We searched for the best match between the mask and the 4$a$ × 4$a$ area around each candidate position of the ellipsoid by calculating the total brightness of (mask × area) for every mask, and adopted the largest value among the 18 total brightnesses. If the largest total brightness was greater than a empirical critical value, it was considered as an ellipsoid. The angular accuracy was only about 10º and the translational position was not very accurate either. Next, our algorithm tried to locate the exact position of the ellipsoid in the 4$a$ × 4$b$ area around the roughly estimated position previously obtained. The area was scanned pixel by pixel and 1° by 1° with a $4a \times 4b$ mask. The total brightness of (mask × area) was calculated at each position and orientation. At the end, a parabolic fit was used to interpolate the position of the maximum brightness and obtain sub-pixel resolution. Figure 1(b) shows that all ellipsoids in the field of view can be captured well with our algorithm. The major tracking error was from the image distortion due to the small tumbling motions of ellipsoids. The orientational resolution was measured to be 1° and the spatial resolution was 0.7 μm along the long axis and 0.3 μm along the short axis from the intercepts of the corresponding translational and rotational *MSD*s at time $t = 0$ [13, 29]. Note that in refs. [13, 14], we used the 2D Gaussian fit, a build-in function in Interactive Data Language (IDL), to identify the position and the orientation of a single ellipsoid. However this fit does not work for multiple ellipsoids in a single frame. The ellipsoid tracking algorithm proposed in ref. [30] can analyze multiple ellipsoids in 3D confocal images, while our algorithm can analyze less clear 2D bright-field images with sub-pixel resolution.

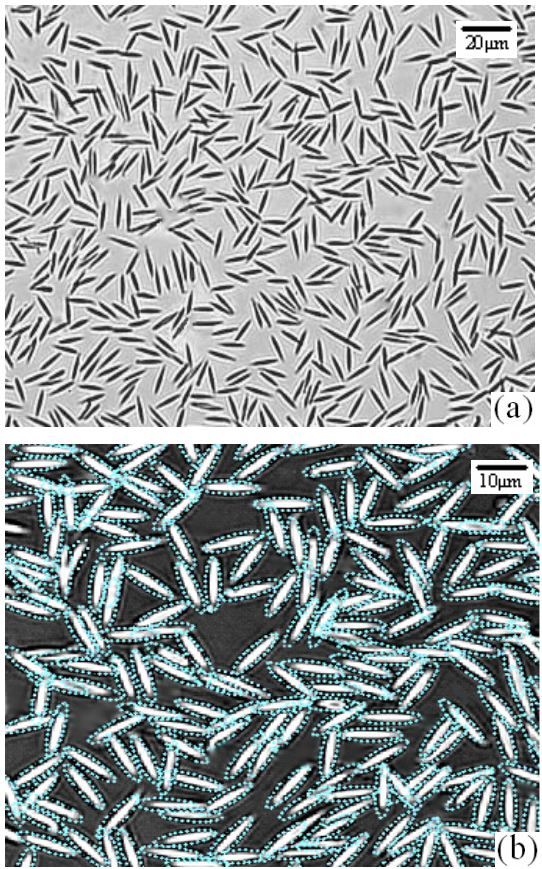

Figure 1. (a) A typical image of a monolayer of ellipsoids with the area fraction of 0.29. (b) A typical particle-tracking result. The dotted blue elliptical contours

represent the positions and orientations of the ellipsoids obtained from image analysis which coincide well with the white images of the real ellipsoids.

### III. Results and Discussions

#### A. Static structures

We measured the static structures and diffusions at 11 different area fractions ranging from 0.01 to 0.68 below the freezing point. The area fraction $\phi = \pi ab\rho$, where $\rho$ is the number density averaged over all frames. The static structure of the monolayer can be characterized by the 2D angular correlation function

$$g_2(r) = \langle \cos(2[\theta_i(0) - \theta_j(r)]) \rangle, \tag{5}$$

where $\theta_i$ is the orientation of the ellipsoid $i$, and $r$ is the center-to-center distance between ellipsoids $i$ and $j$. Figure 2 shows the angular correlation function at different area fractions. All the correlations decay exponentially, which characterizes the short-range orientational order. The quasi-long-range order characterized by algebraic decay was not observed. For the highest concentration in Fig. 2, the correlation decays algebraically at $r/2b < 4$ and decreases more rapidly at larger distances. This is consistent with the observation that the dense suspensions were composed of nematic-like domains with different orientations. The typical domain size is less than $20b$. Within a domain, the ellipsoids are not completely parallel, but form branch-like structures. In contrast, granular spherocylinders with similar aspect ratios can easily form nematic phase in 2D [31], while granular and colloidal rectangles tend to form small nematic domains with tetratic order in 2D [31, 32], These results showed that the microscopic shape of particles can dramatically affect the mesoscopic structure [31]. $\phi = 0.68$ is still in the liquid regime with nematic-like domains and particles can diffuse out of the cages of their neighbors. At further higher area fractions, the monolayer becomes glassy. Simulations have shown that there exists a nematic phase between liquid and solid phases in 2D if the aspect ratio is large ($p \geq 7$ for hard rods [33] and $p \geq 4$ for hard ellipsoids [34]). However our ellipsoids with $p = 9$ changed from liquid to glassy solid with small polycrystal-like nematic domains rather than going through a well-ordered nematic phase. This polycrystal-like nematic structure is similar to the simulation results of the 2D hard rods of $p = 5$ [33] and 2D hard ellipses of $p = 4$ [34]. We attribute the absence of an ordered nematic phase to polydispersity. The 5.8% polydispersity of our ellipsoids induced more disordering in the packing so that the 2D nematic phase shifted to higher aspect ratios.

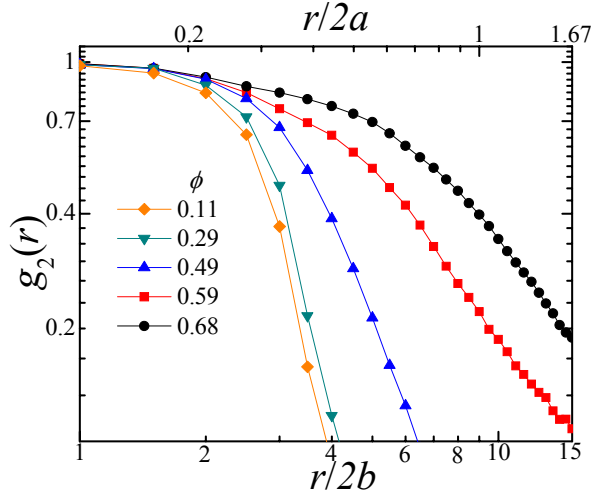

Figure 2. The angular correlation functions $g_2(r)$ of the 2D ellipsoid suspensions at different area fractions. The distance $r$ between two ellipsoids is scaled by $2b$ (bottom axis) and by $2a$ (top axis).

**B. Mean-square displacements (MSDs)**

The dynamic behaviors of ellipsoids in each degree of freedom were characterized by the corresponding $MSD$s. Here we only show $MSD_\parallel$ and $MSD_\perp$ in the body frame because $MSD$ in the lab frame is approximately the sum of them. Note that $MSD_{lab} = \langle \Delta x^2 + \Delta y^2 \rangle = \langle \Delta r^2 \rangle = \langle \Delta r_\parallel^2 + \Delta r_\perp^2 \rangle = MSD_\parallel + MSD_\perp$ at short times. In the long-time regime, the angle of each ellipsoid has been randomized so that $MSD_{lab} = MSD_\parallel + MSD_\perp$ still holds (see ref. [8, 10]). To measure the $MSD$s in the body frame, we first projected the displacements ($\Delta x_n$, $\Delta y_n$) between the $n$'th video frame and ($n$-1)'th video frame in the lab frame onto the body frame ($\Delta a_n$, $\Delta b_n$) [13] by

$$\Delta a_n = \Delta x_n \cos\theta_n + \Delta y_n \sin\theta_n, \tag{6a}$$

$$\Delta b_n = -\Delta x_n \sin\theta_n + \Delta y_n \cos\theta_n, \tag{6b}$$

where $\Delta x_n = x_n - x_{n-1}$. Then we constructed the trajectory of each ellipsoid in its body frame by $a(t) = \sum_n \Delta a_n$ and $b(t) = \sum_n \Delta b_n$. From the trajectories, we calculated *MSD*s along the long and the short axes for each ellipsoid and then obtained the ensemble averaged *MSD*s.

Figures 3-5 show the *MSD*s for the translational motions along the long and short axes in the body frames, and for the rotational motions respectively. The three time regimes can be clearly resolved from the log-log plots in Figs. 3(b), 4(b) and 5(b). In the short- and long-time regimes, the two unit-slope lines on each curve in Figs. 3(b), 4(b) and 5(b) indicate that $\langle \Delta r^2(t) \rangle \sim t$ and the motions are diffusive. In the intermediate time regime, the slopes are less than one so that $\langle \Delta r^2(t) \rangle \sim t^\beta$ with $\beta < 1$ and the motions are sub-diffusive due to the cage effect. Figs. 3(b), 4(b) and 5(b) show that the intermediate time regime starts earlier and ends later at higher concentrations because a higher concentration leads to smaller cages so that particles take less time to hit the neighbors and take more time to diffuse out of the cages. We define the start and end times of the intermediate regime as the point where the local slope deviates by 10% from the unit slope. The measured start and end times of the intermediate regime as a function of area fraction, $\phi$, are shown in Figs. 6(a) and 6(b) respectively. Interestingly both times exhibit power-law relations to $\phi$. Fig. 6(b) can be more quantitatively understood after being re-plotted in Fig. 6(c). Figure 6(c) shows that the start time, $t_b$, has a power-law relation with the "mean free path", $\langle L \rangle = \langle l \rangle - \sqrt{ab}$, where $\langle l \rangle = \sqrt{\pi ab/\phi}$ is the average center-to-center distance between neighboring ellipsoids, $\sqrt{ab}$ is the estimated size of the ellipsoid and $\langle L \rangle$ is the mean distance that particles diffuse before they collide with their neighbors. $t_b \sim \langle L \rangle^{0.75}$ in Fig. 6(c) can be understood from $t_b \sim \langle L \rangle^2 / D_T^S$ where the short-time diffusion coefficient were measured to scale as $D_T^S \sim \langle L \rangle^{1.25}$ (see Fig. 6(d)). The measurement of $D_T^S$ is discussed in section D. These power-law relations are better satisfied at high concentrations where $\langle L \rangle$ is comparable to the cage size.

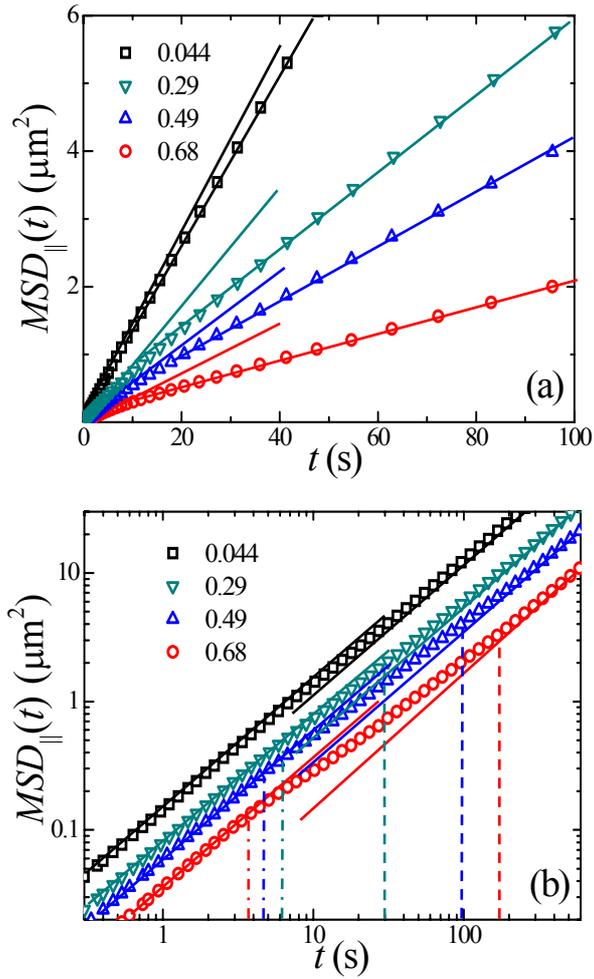

Figure 3. (a) Translational *MSD*s along the long axes of ellipsoids in 2D suspensions with area fraction $\phi$ = 0.044, 0.29, 0.49, 0.68. (b) Log-log plot of (a). The two solid lines with unit slope on each curve in (b) represent the diffusive motions in short- and long-time regimes. The vertical dashed-dotted lines and dashed lines denote the start and end times of the intermediate subdiffusive regime respectively.

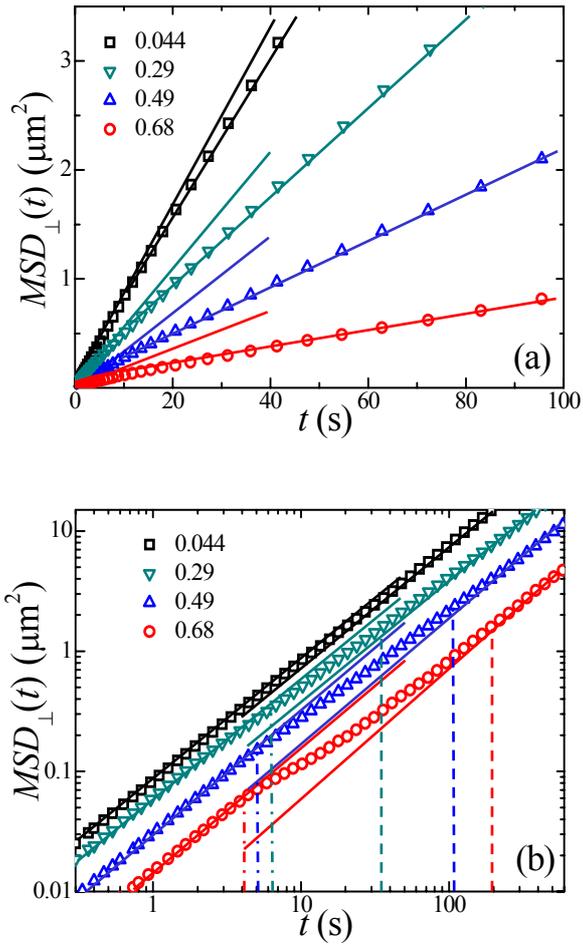

Figure 4. (a) Translational *MSD*s along the short axes of ellipsoids in 2D suspensions with area fraction $\phi$ = 0.044, 0.29, 0.49, 0.68. (b) Log-log plot of (a). The two solid lines with unit slope on each curve in (b) represent the diffusive motions in short- and long-time regimes. The vertical dashed-dotted lines and dashed lines denote the start and end times of the intermediate subdiffusive regime respectively.

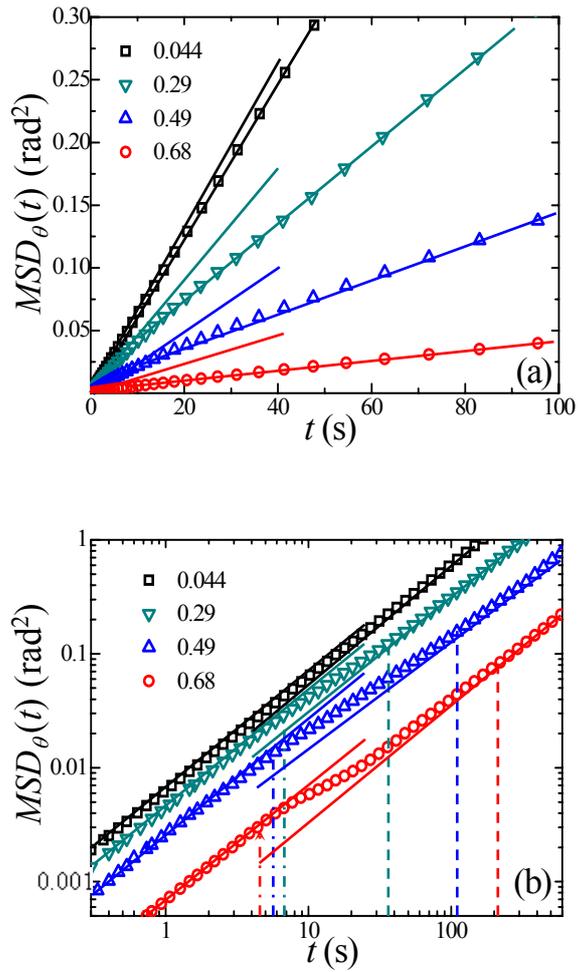

Figure 5. (a) Rotational *MSD*s of ellipsoids in 2D suspensions with area fraction $\phi =$ 0.044, 0.29, 0.49, 0.68. (b) Log-log plot of (a). The two solid lines with unit slope on each curve in (b) represent the diffusive motions in short- and long-time regimes. The vertical dashed-dotted lines and dashed lines denote the start and end times of the intermediate subdiffusive regime respectively.

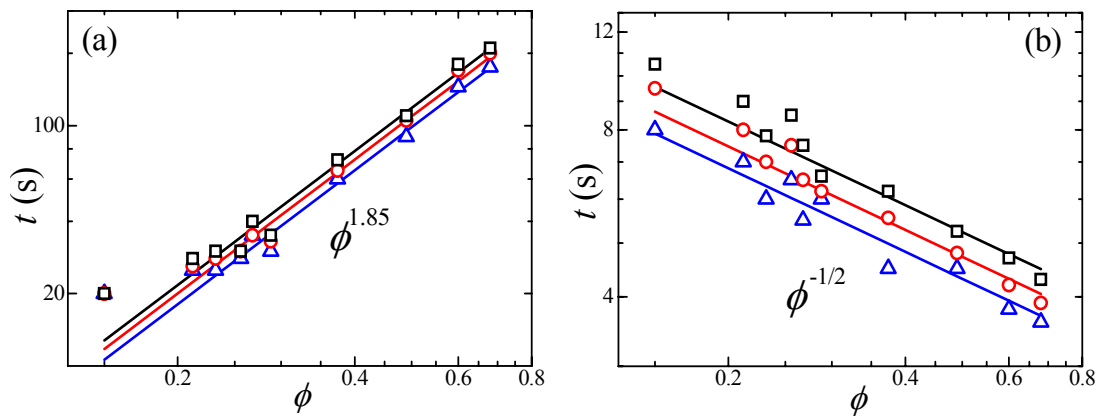

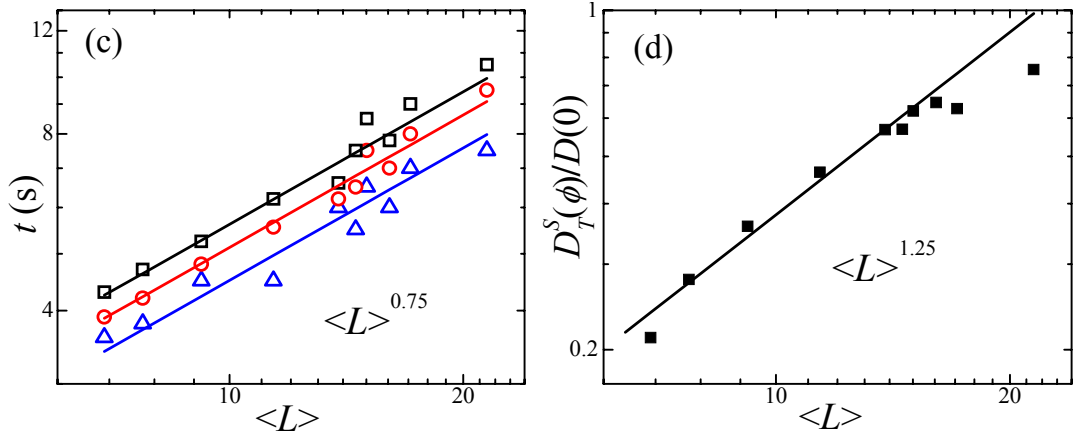

Figure 6. The end time (a) and the start time (b) of the subdiffusive regime at different concentrations. (c) The start time as a function of the "mean free path", $\langle L \rangle$. (d) The normalized short-time translational diffusion coefficient, $D_T^S(\phi)/D_T^S(0)$, from Fig. 8 in section D. The symbols △ and ○ denote translational motions along the long and the short axes respectively; The symbol □ denotes rotational motions.

## C. Long-time translational and rotational diffusion coefficients

The long-time translational and rotational diffusion coefficients were obtained from the slopes of the corresponding MSDs. Figure 7 shows the normalized diffusion coefficient $D^L(\phi)/D(0)$. All ratios decay from 1 at infinite dilution to less than 0.1 at $\phi = 0.68$. At low densities, the translational motions along the long axis are more easily caged than the translational motions along the short axis or the rotational motions. Consequently $D_\parallel^L(\phi)/D_\parallel(0)$ decays more quickly than $D_\perp^L(\phi)/D_\perp(0)$ and $D_\theta^L(\phi)/D_\theta(0)$ as shown in Fig. 7. At high concentrations, $D_\parallel^L(\phi)/D_\parallel(0)$ and $D_\perp^L(\phi)/D_\perp(0)$ have the same values, and $D_\theta^L(\phi)/D_\theta(0)$ is smaller. Hence the long-time rotational diffusion is more strongly caged than the translational diffusion at high concentrations. $D_T^L(\phi)/D_T(\phi)$ is between the $D_\parallel^L(\phi)/D_\parallel(0)$ and $D_\perp^L(\phi)/D_\perp(0)$ curves as expected. Although there is no theoretical or simulation results on 2D ellipsoid diffusion to compare with, our $D_{T,\theta}^L(\phi)/D_{T,\theta}(0)$ have similar shapes as those in the simulation of 2D rods [35] except that our translational diffusion coefficient decays more rapidly in the dilute regime. This can be interpreted from the following two aspects: (1) Our ellipsoids have larger aspect ratio ($p = 9$) than the rods in ref. [35] ($p = 6$) so that the cage effect along the long axis is stronger at low concentrations. This behavior has also been found in the simulation of rods in 3D suspensions [36]. (2) The Brownian dynamics simulations neglect the hydrodynamic interactions which will slightly reduce the diffusion coefficients [36, 37]. Compared with the simulation results of rod diffusions in 3D [36], our $D_{T,\theta}^L(\phi)/D_{T,\theta}(0)$ decreases more slowly due to the presence of the bottom wall. The wall can be approximately regarded as dense ellipsoids, so that the whole area fraction range in 2D corresponds only to the intermediate volume fraction range in 3D. Consequently, $D_{T,\theta}^L(\phi)/D_{T,\theta}(0)$ for ellipsoids in 2D changes less dramatically than that for rods in 3D.

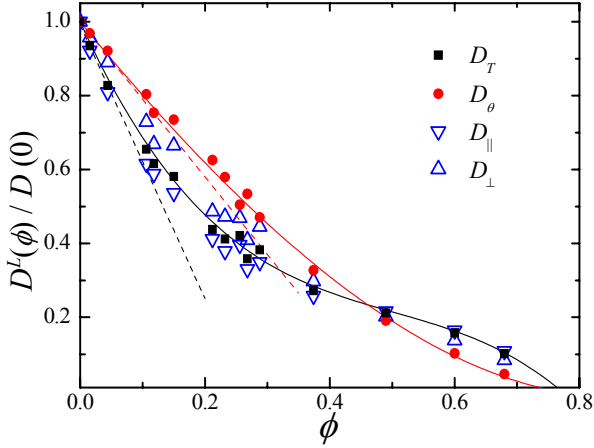

Figure 7. The normalized long-time diffusion coefficients $D^L(\phi)/D(0)$ decrease with the area fraction $\phi$. The symbol ▽ represents $D_\parallel^L(\phi)/D_\parallel(0)$; The symbol △ represents $D_\perp^L(\phi)/D_\perp(0)$; The symbol ■ represents $D_T^L(\phi)/D_T(0)$ in the lab frame; The symbol ● represents $D_\theta^L(\phi)/D_\theta(0)$. $D(0)$ is the diffusion coefficient at the infinite dilute limit. The solid curves are the 3rd-order polynomial fits to $D_T^L(\phi)/D_T(0)$ and $D_\theta^L(\phi)/D_\theta(0)$. The dashed lines show the first-order linear terms in the polynomials.

The concentration dependence of the diffusion coefficient is often empirically expressed by a polynomial because the analytic form is not available. The diffusion of colloidal spheres follows the Batchelor's formulae: $D_T^L(\phi)/D_T(0) = 1 - 2.1\phi + O(\phi^2)$ at long-time limit [38] and $D_T^S(\phi)/D_T(0) = 1 - 1.83\phi + O(\phi^2)$ at short-time limit [39] in 3D suspensions. In 2D suspensions, the theoretical prediction is $D_T^L(\phi)/D_T(0) = 1 - 2\phi + O(\phi^2)$ for hard disks at the long-time limit [40]. In experiment, $D_T^S(\phi)/D_T^S(0)$ for a colloidal monolayer of hard spheres has been measured by the evanescent wave dynamic light scattering [28]; it decays more slowly than that of a 3D suspension because the presence of the wall largely screen out particles' hydrodynamic interaction and weaken the drag force of neighbors. Our measured diffusion coefficients in Fig. 7 can be well fitted by 3rd-order polynomials:

$$D_T^{L,S}(\phi)/D_T(0) = 1 + \alpha_{T1}^{L,S}\phi + \alpha_{T2}^{L,S}\phi^2 + \alpha_{T3}^{L,S}\phi^3, \quad (7a)$$

$$D_\theta^{L,S}(\phi)/D_\theta(0) = 1 + \alpha_{\theta1}^{L,S}\phi + \alpha_{\theta2}^{L,S}\phi^2 + \alpha_{\theta3}^{L,S}\phi^3. \quad (7b)$$

The superscript $S$ represents the short-time diffusions shown in Fig. 8 in the next section. The fitting coefficients are listed in Table 1. The first-order and the second-order coefficients have opposite signs for both long-time translational and rotational diffusions, which agrees with the behavior of hard rods observed in the simulation in ref. [36] (see Table 1). The coefficients of hard rods in 2D in Table 1 were extrapolated from the simulation results of $p > 6$ ellipsoids in ref. [36]. The $D_T^L(\phi)/D_T^L(0)$ of ellipsoids decays more quickly than that of spheres in ref. [40]. This is consistent with the rod diffusions in 3D where the first-order coefficient decreases with aspect ratio $p$ [36].

Table 1. The polynomial fitting coefficients of $D_T^{L,S}(\phi)/D_T(0)$ and $D_\theta^{L,S}(\phi)/D_\theta(0)$ in Eq. (7). The fitting curves are shown in Fig. 7 and Fig. 8. Columns 2 to 5: coefficients for our ellipsoids with $p = 9$ in 2D. Columns 6 and 7: coefficients for rods with $p = 9$ in the 2D simulation [36]. Column 8: theoretical first-order coefficients $\alpha = -2 - {}^{10}/_{32}(p-1) - {}^{1}/_{53}(p-1)^2$ for rods with $p = 9$ in 3D [41].

| Poly-nomial order | $\alpha_T^L$ (ellipsoid) | $\alpha_\theta^L$ (ellipsoid) | $\alpha_T^S$ (ellipsoid) | $\alpha_\theta^S$ (ellipsoid) | $\alpha_T^L$ (rod [36]) | $\alpha_\theta^L$ (rod [36]) | $\alpha_T^L$ (rod [41]) |
|---|---|---|---|---|---|---|---|
| 1 | -3.74 ± 0.2 | -2.06 ± 0.1 | -2.1 ± 0.1 | -0.94 ± 0.05 | -8.51 | -3.03 | -5.71 |
| 2 | 6.50 ± 0.4 | 0.68 ± 0.03 | 2.4 ± 0.2 | -0.915 ± 0.07 | 24.30 | 39.45 | |
| 3 | -4.32 ± 0.4 | 0.40 ± 0.03 | -1.54 ± 0.15 | 0.65 ± 0.06 | -21.08 | -100.8 | |

The extrapolations in Fig. 7 show that the long-time translational motions are completely caged near $\phi = 0.77$, which is very close to the lower limit of 0.76 for $p = 6$ ellipsoids in 2D [34] and 0.78 for $p = 6$ rods in 2D [33] at the liquid-solid transition. In contrast, the rotational diffusions in Fig. 7 are completely caged at the lower concentration of $\phi = 0.73$. The long-time translational motions caged before rotational motions have also been observed in the simulations of rods in 2D [35] and in 3D [36]. At $\phi = 0.68$, all ellipsoids in one domain collectively rotated very slowly at about 1 rad/hour, and some ellipsoids could diffuse out of the domains. We also qualitatively observed a higher concentration at $\phi \sim 0.74$ from the video, where ellipsoids could still diffuse out of cages, but the collective rotations of the domains were frozen.

**D. Short-time translational and rotational diffusion coefficients**

The short-time diffusions are mainly affected by the complex hydrodynamic interactions between particles. The measurements of short-time diffusions of anisotropic particles are rather limited. Maeda *et al.* have reported the short-time diffusions of rods in the bottom layers of concentrated 3D samples [42]. Here we measured the short-time diffusions in monolayers of ellipsoids (see Fig. 8). Their concentration dependence is similar to the long-time diffusion coefficients shown in Fig. 7, but with higher values because the cage effect mainly reduces long-time diffusions. $D_\parallel^S(\phi)/D_\parallel(0)$ in Fig. 8 decays more rapidly than $D_\perp^S(\phi)/D_\perp(0)$ and $D_\theta^S(\phi)/D_\theta(0)$ at low area fraction because the translational motions along the long axis is more easily caged than the translational motions along short axis or the rotational motions. The polynomial fitting coefficients are listed in Table.1. $D_T^S(\phi)/D_T(0)$ for our ellipsoids decays faster than that for a monolayer of spheres near one wall ($\alpha_{T1} > -1.4$) [28].

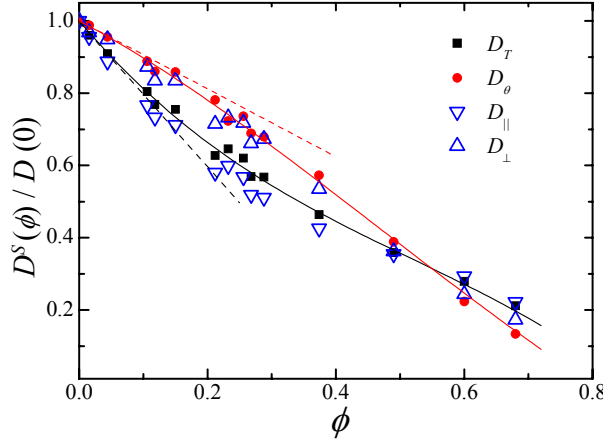

Figure 8. The normalized short-time diffusion coefficient $D^S(\phi)/D(0)$ decreases with the area fraction $\phi$. The symbol $\triangledown$ represents $D_\parallel^S(\phi)/D_\parallel(0)$; The symbol $\triangle$ represents $D_\perp^S(\phi)/D_\perp(0)$; The symbol ■ represents $D_T^S(\phi)/D_T(0)$ in the lab frame; The symbol ● represents $D_\theta^S(\phi)/D_\theta(0)$. $D(0)$ is the corresponding diffusion coefficient at the infinite dilute limit. The solid curves are the 3rd-order polynomial fits to $D_T^S(\phi)/D_T(0)$ and $D_\theta^S(\phi)/D_\theta(0)$. The dashed lines show the first-order linear terms in the polynomials.

**E. Diffusion anisotropy**

The anisotropic particle shape leads to the anisotropic drag coefficient and anisotropic diffusions. For an ellipsoid, the diffusion anisotropy can be characterized by the ratio of the diffusion coefficients along the long and the short axes (see Fig. 9).

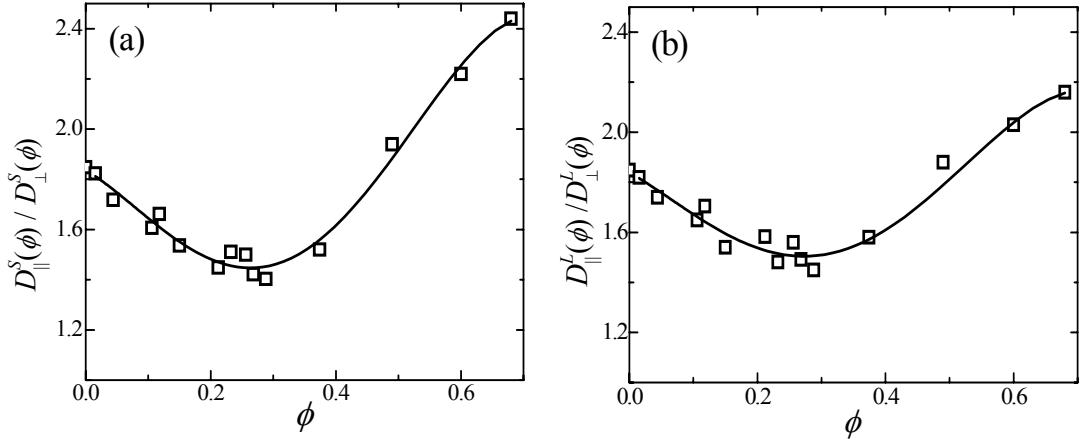

Figure 9. The diffusion anisotropy $D_{\parallel}(\phi)/D_{\perp}(\phi)$ at (a) short-time limit; (b) long-time limit. The solid curves are guide for eyes.

At both short- and long-time limits, the ratios exhibit minima at $\phi \sim 0.3$. This indicates that the motion along the long axis is more caged at $\phi < 0.3$, while the motion along the short axis becomes more and more caged at $\phi > 0.3$. Fig. 1(a) shows the configuration of ellipsoids at $\phi \approx 0.3$. When $\phi > 0.6$, the local nematic structures provide additional anisotropy so that the ratios are larger than two, which breaks the upper bound of $D_{\parallel}(0)/D_{\perp}(0) = 2$ in 3D, see Eq. (8) derived from Eqs. (1-3) at $p \to \infty$ limit [23]:

$$D_{\parallel} = \frac{k_B T \ln p}{2\pi \eta a}, \tag{8a}$$

$$D_{\perp} = \frac{k_B T \ln p}{4\pi \eta a}. \tag{8b}$$

At the dilute limit, we measured $D_{\parallel}(0) = 0.079$ μm²/s, $D_{\perp}(0) = 0.043$ μm²/s and $D_\theta(0) = 0.0033$ rad²/s by averaging the diffusion coefficients of 15 isolated ellipsoids with neighbor distance $> 100a$. According to Eqs. (1-3), the corresponding diffusion coefficients of the same-sized ellipsoid in the 3D dilute limit would be 0.149 μm²/s, 0.104 μm²/s and 0.00625 rad²/s respectively. The measured 2D diffusion coefficients are about half of those values in 3D due to the friction exerted by the wall. The ratio $D_{\parallel}(0)/D_{\perp}(0)$ is raised from 1.43 in 3D to 1.84 in 2D, which indicates that the drag imposed by the wall is larger along the short axis than along the long axis. This wall-induced diffusion anisotropy can be qualitatively explained as in the following. For a horizontal ellipsoid in 3D, the horizontal fluid flow along the short axis tends to go around from top and bottom with 2*b* displacement, while the flow along the long axis goes around the ellipsoid by 2*b* displacement from top, bottom left and right. Adding a bottom wall will largely block the pathway at the bottom so that $D_\perp$ is affected by ~50% while $D_\parallel$ is only affected by ~25%. This effect can induce a stronger diffusion anisotropy for ellipsoids confined between two walls as reported in ref. [14].

**F. Effective diffusion coefficients**

The nonlinear MSDs in Figs. 3-5 can be re-plotted as linear functions as shown in Figs. 10 and 11. Their y axes are the *effective* translational and rotational diffusion coefficients defined as $\tilde{D}(t) = MSD/(2t)$, which is slightly different from the diffusion coefficient $D(t) = \Delta MSD/(2\Delta t)$. The effective diffusion coefficient is directly related to the relaxation function $\mu(t) = D - D^L$ which is the difference between the self-diffusion coefficient $D$ and its dominating term $D^L$. For 2D hard disks, $\mu(t)$ is found to decay according to $1/t$ during $\tau_H < t < \tau_D$ by solving the Smoluchowski equation [40]. For micrometer-sized particles, $\tau_H \sim$ ms is the time after which the hydrodynamic effects become important and $\tau_D \sim$ s is the time for a particle to diffuse a hydrodynamic screening length [40, 43]. The time integration of $\mu(t)$ leads to [44]

$$MSD/2 = D^L t + \int_0^t \mu(t')dt' \approx D^L t + (D^S - D^L)\tau_L \ln(t/\tau_M) + o(1), \qquad (9)$$

where $\tau_L$ and $\tau_M$ are the long- and medium-scale time constants respectively and depend on the nature of interactions and concentrations [43-45]; and $D^S$ is the short-time self-diffusion coefficient. The logarithm term in Eq. (9) reflects the fact that the asymptotic value of the diffusion coefficient is reached very slowly. Eq. (10) indicates that the effective translational diffusion coefficient

$$\tilde{D}(t) = D^L + (D^S - D^L)\tau_L (\ln(t/\tau_M))/t + o(1), \qquad (10)$$

which decays linearly with $(\ln t)/t$ from the initial value $D_T^S$ to the long-time asymptotic value $D_T^L$ in 2D Brownian suspension. This relation for hard disks has been confirmed experimentally in monolayers of colloidal spheres [43, 45, 46]. Here we observed that Eq. (10) still holds for ellipsoids' translational diffusions along both long and short axes and rotational diffusions at all concentrations (see Figs. 10 and 11)

This suggests that the diffusion of ellipsoids along the long and short axes is dominated by binary collisions and that the mode coupling effect is weak at $\phi < 0.68$ [46].

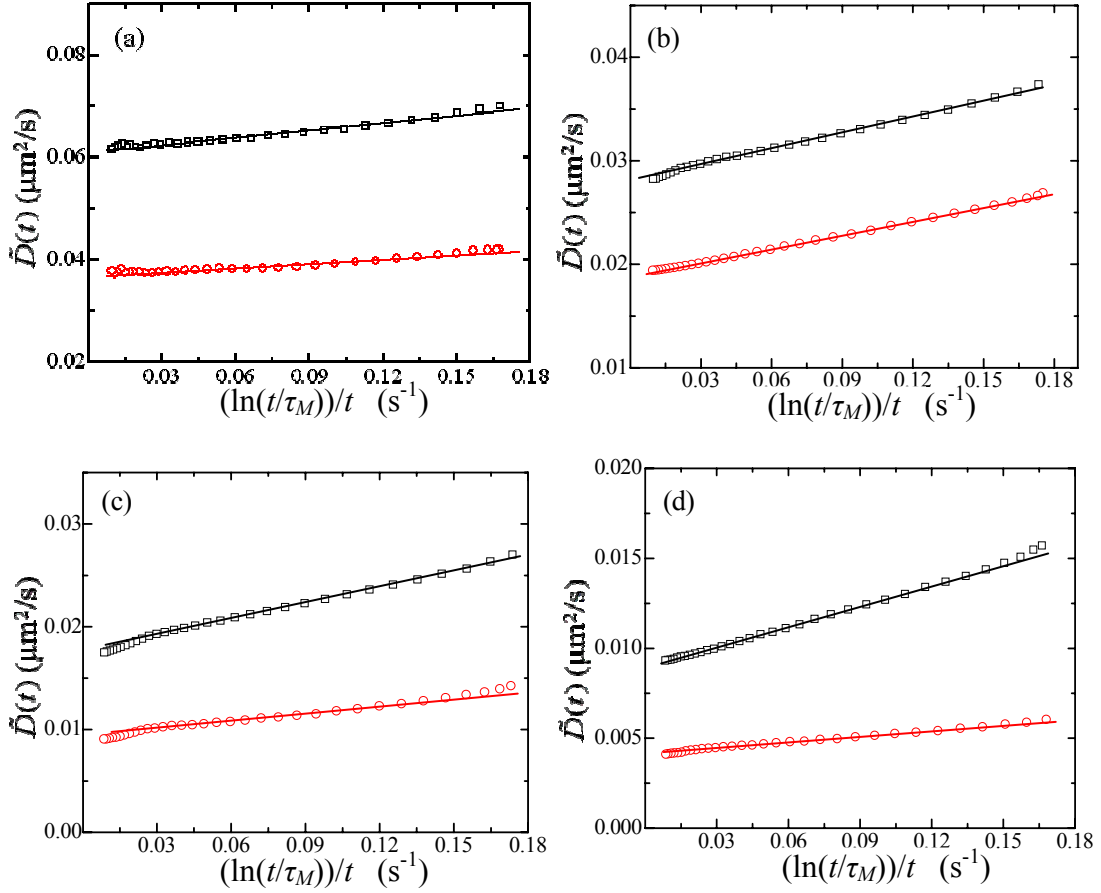

Figure 10. The effective translational diffusion coefficient $\tilde{D}(t)$ verses $(\ln(t/\tau_M))/t$ at different area fractions (a) $\phi = 0.044$, (b) 0.29, (c) 0.49 and (d) 0.68. The time ranges from 5 s to 600 s. The symbols □ and ○ denote translation diffusions along the long and short axes respectively. The solid lines are linear fits. The corresponding $\tau_M$'s fitted from Eq. (10) are listed in Table 2.

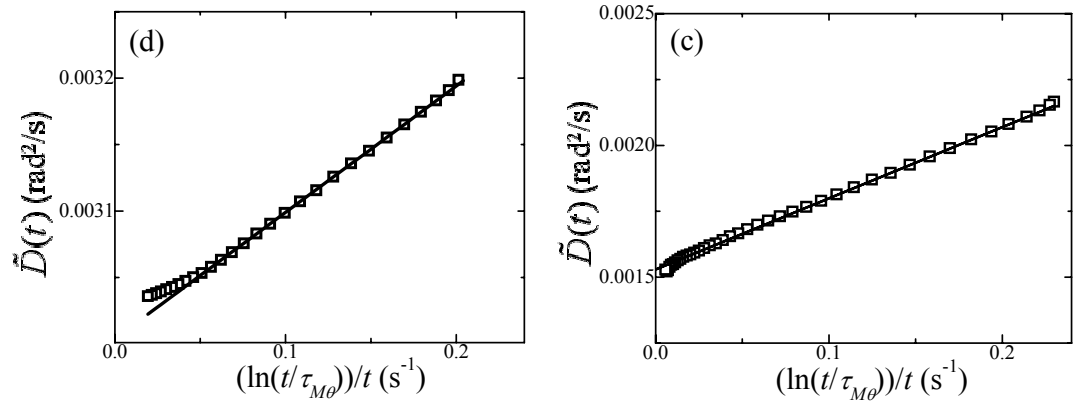

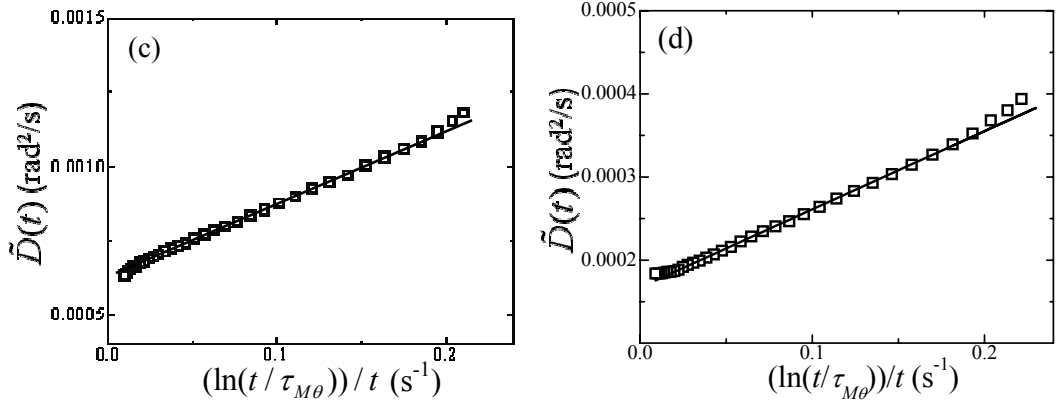

Figure 11. The effective rotational diffusion coefficient $\tilde{D}(t)$ verses $(\ln(t/\tau_M))/t$ at different area fractions (a) $\phi = 0.044$, (b) 0.29, (c) 0.49 and (d) 0.68. The time ranges from 5 s to 600 s. The solid lines are linear fits. The corresponding $\tau_{M\theta}$'s fitted from Eq. (10) are listed in Table 2.

The values $D^L_{\parallel,\perp,\theta}(t)$ and $D^S_{\parallel,\perp,\theta}$ were directly measured from the slopes of the $MSD(t)$ and shown in Figs. 7 and 8. Given $D^L_{\parallel,\perp,\theta}(t)$ and $D^S_{\parallel,\perp,\theta}$, $\tau_L$ and $\tau_M$ can be fitted from Eq. (10). Their values are listed in Table 2. For each degree of freedom, $\tau_M$ is almost a constant while $\tau_L$ decreases with the area fraction. These behaviors of $\tau_L$ and $\tau_M$ have also been observed in the diffusion of colloidal spheres [45].

Table 2. The fitted values of the time constants $\tau_L$ and $\tau_M$ from Eq. (10) for translational motions along the long ($\parallel$) and short ($\perp$) axes and for rotational motions ($\theta$).

| $\phi$ | $\tau_{L\parallel}$ (s) | $\tau_{L\perp}$ (s) | $\tau_{L\theta}$ (s) | $\tau_{M\parallel}$ (s) | $\tau_{M\perp}$ (s) | $\tau_{M\theta}$ (s) |
|---|---|---|---|---|---|---|
| 0.044 | 8.7 | 12 | 7.3 | 2.0 | 2.2 | 1.8 |
| 0.29 | 6.9 | 7.8 | 4.5 | 1.9 | 2.1 | 1.6 |
| 0.49 | 6.2 | 6.2 | 4.1 | 1.9 | 2.1 | 1.7 |
| 0.68 | 4.8 | 4.9 | 3.4 | 2.2 | 2.0 | 1.6 |

### G. Anisotropic to isotropic diffusion

The drag coefficients $\gamma_\parallel$ and $\gamma_\perp$ in the body frame are constant, but $\gamma_x$ and $\gamma_y$ in the lab frame depends on the ellipsoid's orientation which changes with time. Consequently the translational motion along the $x$ or $y$ direction in the lab frame is coupled with the rotational motion [13]. Such translation-rotation coupling introduces substantial complications in the Langevin or the Fokker-Planck equation. These complications are frequently circumvented by assuming that the initial anisotropic diffusion lasts only for very short times and then the isotropic diffusion will be recovered at practical time scales. Ref. [13] showed that the transient from anisotropic to isotropic diffusion of a micrometer-sized particle is about a few seconds which corresponds to the angular diffusion time $\tau_\theta = 1/(2D_\theta^L)$ for an ellipsoid to diffuse 1 rad. At longer times, the orientational angle becomes random and the translational drag coefficient becomes a constant after averaging over all angles. Consequently the translational motion is decoupled from the rotational motion at $t \gg \tau_\theta$ and the ellipsoid can be treated as a sphere. Here we measured the crossover from the initial anisotropic diffusion to the long-time isotropic diffusion caused by the translation-rotation coupling. The $x$ axis in the lab frame is defined as the initial orientation of the ellipsoid. The diffusion coefficient along the $x$ and $y$ directions are $D_{xx}(t) = \langle \Delta x^2(t) \rangle / (2t)$ and $D_{yy}(t) = \langle \Delta y^2(t) \rangle / (2t)$, respectively. The diffusion is anisotropic ($D_{xx}(t) > D_{yy}(t)$) at $t < \tau_\theta$ and becomes isotropic ($D_{xx}(t) = D_{yy}(t)$) at $t \gg \tau_\theta$ when directional memory is washed out. The crossover has been measured in the experiment of single ellipsoid in 2D [13] and in the simulation of concentrated rods in 2D [47]. Here we measured $D_{xx}(t)$ and $D_{yy}(t)$ normalized by $(D_{ii} - \bar{D})/\Delta D$, where $\bar{D} = (D_\parallel^L + D_\perp^L)/2$ and $\Delta D = (D_\parallel^L - D_\perp^L)/2$ (see Fig. 11). With such normalization, $D_{xx}(t)$ or $D_{yy}(t)$ should change from +1 or -1 to 0 at the dilute limit [13]. In Fig. 11 $\phi$ = 0.01, $D_{xx}(t)$ or $D_{yy}(t)$ changes from +1 or -1 to 0 at the time scale $\tau_\theta$ which agrees with the single ellipsoid measurement in ref. [13]. At higher concentrations, however, each curve in Fig. 12 changes dramatically at two time scales. The first time scale, $\tau_\theta$, for $\phi$ = 0.01, 0.15, 0.29 and 0.60 in Fig. 11 agrees well with the measured $\tau_\theta$ = 150 s, 200 s, 300 s and 1500 s from the corresponding rotational *MSD*s in Fig. 7. For $\phi$ = 0.60, our experimental time was not long enough for us to observe the isotropic diffusion with good statistics. The second time scale is at ~ 10 s which characterizes the time for ellipsoids to diffuse to their neighbors so that the diffusion coefficients would decrease dramatically. This time scale agrees well with the start time of the intermediate nondiffusive time regime shown in Fig.

6(a). The two decays in each $D_{xx}(t)$ or $D_{yy}(t)$ curve have also been observed in the simulation of rods at finite concentrations in 2D [47]. At $t < 10$ s, $D_{xx}$ increases while $D_{yy}$ decreases when $\phi$ changes from 0.29 to 0.60. This stronger diffusion anisotropy at higher area fraction reflects the local nematic structures which provide highly anisotropic cages.

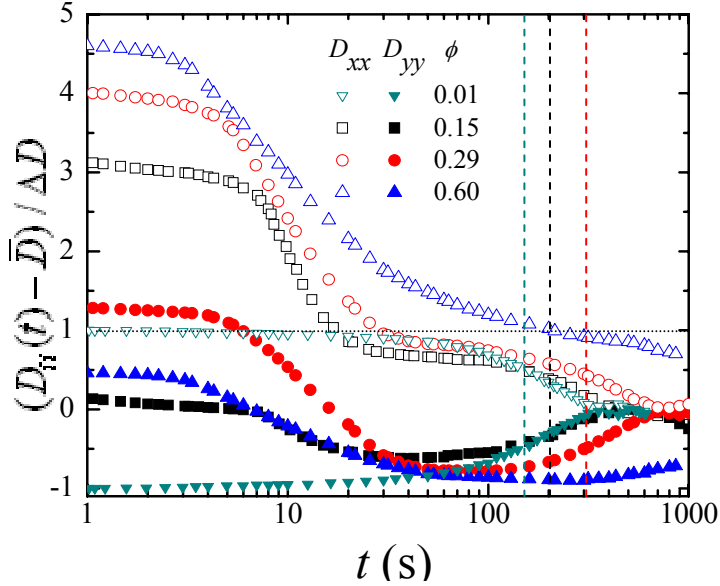

Figure 12. The normalized diffusion coefficients $(D_{xx} - \bar{D})/\Delta D$ (open symbols) and $(D_{yy} - \bar{D})/\Delta D$ (solid symbols) in the lab frame. The $x$ axis is chosen to be along the long axis of the initial position of each ellipsoid. The vertical dashed lines indicate the $\tau_\theta$ for different densities. The $\tau_\theta$ for $\phi = 0.60$ is beyond the time window.

## IV. Summary

We have developed an algorithm to track ellipsoids in 2D with sub-pixel resolution. The ellipsoid diffusions in monolayers near a wall were measured at 11 area fractions ranging from 0.01 to 0.68. The translational and rotational motions are diffusive in the short-time and long-time regimes, and become sub-diffusive in the intermediate time regime. Both the start and end times of the intermediate regime exhibit power-law scaling to the area fraction. The long-time and short-time diffusion coefficients of both translational and rotational motions, $(D_T^L, D_\theta^L, D_T^S, D_\theta^S)$, were fitted by third-order polynomials of area fraction. As the area fraction increased from 0 to 0.3, the diffusion along the long axis decreased more quickly than the diffusion along the short axis or the rotational diffusion. This reflects the fact that the translational motion along the long axis is more easily caged. Hence, both long-time and short-time $D_\parallel(\phi)/D_\perp(\phi)$ reach the minima at the intermediate area fraction (~30%) where the motions are mainly caged along long axes, but not along short axes. The wall further enhances the diffusion anisotropy characterized by the ratio $D_\parallel(\phi)/D_\perp(\phi)$. The coupling effect between translational and rotational motions was elucidated by the crossing over from anisotropic to isotropic diffusion. The cage effect and a particle's anisotropic shape are responsible for the two observed time scales in the lab-frame diffusion coefficients, $D_{xx}(t)$ and $D_{yy}(t)$.

For the 2D suspension of hard spheres, the effective translational diffusion coefficient is linearly related to (ln$t$)/$t$ when the diffusions are dominated by binary collisions and the mode coupling effect is weak [43-46]. Here we observed that this behavior still holds in 2D ellipsoid diffusions, including rotational and translational motions along the long and short axes for area fractions ranging from 0.01 to 0.68. This result suggests that the relaxation functions $\mu(t)$ of both the translational and rotational motions decay according to 1/$t$, which provides new challenges in theory.

## ACKNOWLEDGEMENTS

This work was supported by the Hong Kong University of Science and Technology grant RPC07/08.SC04 and the William Mong Institute of Nano Science and Technology at Hong Kong University of Science and Technology.**References:**
1. B. P. Binks, Curr. Opin. Colloid Interface Sci. **7**, 21 (2002).
2. M. Fuji, H. Fujimori, T. Takei, T. Watanabe, and M. Chikazawa, J. Phys. Chem. B **102**, 10498 (1998).
3. H. Jia, G. Zhu, and P. Wang, Biotechnol. Bioeng. **84**, 406 (2003).
4. R. Aveyrad, B. P. Pinks, and J. H. Clint, Adv. Colloid Interface Sci. **100**, 503 (2003).
5. P. G. Saffman and M. Delbruck, Proc. Natl. Acad. Sci. U.S.A. **72**, 3111 (1975); A. Sokolov, I. S. Aranson, J. O. Kessler, and R. E. Goldstein, Phys. Rev. Lett. **98**, 158102 (2007).
6. R. Cush, P. S. Russo, Z. Kucukyavuz, Z. Bu, D. Neau, D. Shih, S. Kucukyavuz, and H. Ricks, Macromolecules **30**, 4920 (1997).
7. P. S. Russo, M. Baylis, Z. Bu, W. Stryjewski, G. Doucet, E. Temyanko, and D. Tipton, J. Chem. Phys. **111**, 1746 (1999).

8. M. P. B. van Bruggen, H. N. W. Lekkerkerker, and J. K. G. Dhont, Phys. Rev. E **56**, 4394 (1997); M. P. B. van Bruggen, H. N. W. Lekkerkerker, G. Maret, and J. K. G. Dhont, Phys. Rev. E **58**, 7668 (1998).
9. H. Krammer, M. Deggelman, C. Graf, M. Hagenbüchle, C. Johner, and R. Weber, Macromolecules **25**, 4325 (1992).
10. J. K. Phalakornkul, A. P. Gast, and R. Pecora, Macromolecules **32**, 3122 (1999).
11. C. Lellig, J. Wagner, R. Hempelmann, S. Keller, D. Lumma, and W. Härtl, J. Chem. Phys. **121**, 7022 (2004).
12. A. Wilk, J. Gapinski, A. Patkowski, and R. Pecora, J. Chem. Phys. **121**, 10794 (2004).
13. Y. Han, A. M. Alsayed, M. Nobili, J. Zhang, T. C. Lubensky, and A. G. Yodh, Science **314**, 626 (2006).
14. Y. Han, A. Alsayed, M. Nobili, and A. G. Yodh, Phys. Rev. E **80**, 6 (2009).
15. R. Duggal and M. Pasquali, Phys. Rev. Lett. **96**, 246104 (2006).
16. B Bhaduri, A Neild, and TW Ng, Appl. Phys. Lett. **92**, 084105 (2008).
17. G. Li and J. X. Tang, Phys. Rev. E **69**, 061921 (2004).
18. B. J. Berne and R. Pecora, *Dynamic Light Scattering* (Dover, New York, 2000).
19. D. W. Schaefer, G. B. Benedek, P. Schofield, and E. Bradford, J. Chem. Phys. **55**, 3884 (1971).
20. W. Eimer, J. R. Williamson, S. G. Boxer, and R. Pecora, Biochemistry **29**, 799 (1990).
21. Y. Yin, C. Zhao, S. Kuroki, and I. Ando, Macromolecules **35**, 2335 (2002).
22. F. C. Cheong and D. G. Grier, Opt. Express **18**, 6555 (2010).
23. J. Happel and H. Brenner, *Low Reynolds Number Hydrodynamics* (Kluwer, Dordrecht, 1991).
24. F. Perrin, J. Phys. Radium **5**, 497 (1934).
25. S. Koenig, Biopolymers **14**, 2421 (1975).
26. P. N. Segrè, O. P. Behrend, and P. N. Pusey, Phys. Rev. E 52, 5070 (1995).
27. C. C. Ho, A. Keller, J. A. Odell, and R. H. Ottewill, Colloid Polym Sci. **271**, 469 (1993).
28. V. N. Michailidou, G. Petekidis, J. W. Swan, and J. F. Brady, Phys. Rev. Lett. **102**, 068302 (2009).
29. J. C. Crocker and D. G. Grier, J. Colloid Interface Sci. **179**, 298 (1996).
30. A. Mohraz and M. J. Solomon, Langmuir **21,** 5298 (2005).
31. V. Narayam, N. Menon, and S. Ramaswamy, J Stat Mech: Theory Exp **2006**, P01005 (2006).
32. K. Zhao, C. Harrison, D. Huse, W.B. Russel, and P.M. Chaikin, Phys. Rev. E **76**, 040401(R) (2007).
33. M. A. Bates and D. Frenkel, J. Chem. Phys. **112**, 10034 (2000).
34. J. A. Cuesta and D. Frenkel, Phys. Rev. A **42**, 2126 (1990).
35. J. M. Lahtinen, T. Hjelt, T. Ala-Nissila, and Z. Chvoj, Phys. Rev. E **64**, 021204 (2001).
36. H. Löwen, Phys. Rev. E **50**, 1232 (1994).
37. V. Pryamitsyn and V. Ganesan, J. Chem. Phys. **128**, 134901 (2008).
38. G. K. Batchelor, J. Fluid Mech. **131**, 155 (1983); Cichocki and B. U. Felgerhof, J. Chem. Phys. **89**, 3705 (1988).
39. G. K. Batchelor, J. Fluid Mech. **74**, 1 (1976).
40. B. J. Ackerson and L. Fleishman, J. Chem. Phys. **76**, 2675 (1982).
41. J. K. G. Dhont, M. P. B. van Bruggen, and W. J. Briels, Macromolecules **32**, 3809 (1999).
42. H. Maeda and Y. Maeda, Nano lett. **7**, 3329 (2007).

[43] A. H. Marcus, B. H. Lin, and S. A. Rice, Phys. Rev. E **53**, 1765 (1996).
[44] B. Cichocki and B. U. Felderhof, J. Phys. Condens. Matter **6**, 7287 (1994).
[45] A. H. Marcus, J. Schofield and S. A. Rice, Phys. Rev. E **60**, 5725 (1999).
[46] J. Schodield, A. H. Marcus, and S. A. Rice, J. Phys. Chem. **100**, 18950 (1996).
[47] T. Munk, F. Höfling, E. Frey, and T. Franosch, Europhys. Lett. **85**, 30003 (2009).